\documentstyle[epsfig]{texas}

\begin{document}

\title{EVOLUTIONARY SEQUENCES OF IRROTATIONAL BINARY NEUTRON STARS}
\author{Silvano Bonazzola, Eric Gourgoulhon and Jean-Alain Marck}
\address{D\'epartement d'Astrophysique Relativiste et de Cosmologie \\
  UMR 8629 du C.N.R.S., Observatoire de Paris, \\
  F-92195 Meudon Cedex, France \\
{\rm Email: Silvano.Bonazzola, Eric.Gourgoulhon, Jean-Alain.Marck@obspm.fr}}

\begin{abstract}
We present results of numerical computations of quasiequilibrium sequences
of binary neutron stars with zero vorticity, in the general relativistic
framework. The Einstein equations are solved under the assumption
of a conformally flat spatial 3-metric (Wilson-Mathews approximation). The
evolution of the central density of each star is monitored as the orbit
shrinks in response to gravitational wave emission. 
For a compactification ratio $M/R=0.14$, the central density 
remains rather constant (with a slight increase, below $0.1\%$)
before decreasing. For a higher compactification ratio $M/R=0.17$ (i.e. stars
closer to the maximum mass configuration), a very small density increase
(at most $0.3\%$) is observed before the decrease. 
This effect remains within the error induced by the
conformally flat approximation. 
It can be thus concluded that no substantial compression of the stars is found,
which would have
indicated a tendency to individually collapse to black hole prior to merger. 
Moreover, no turning point has been found in the binding energy or angular
momentum along evolutionary sequences, which may indicate that these
systems do not have any innermost stable circular orbit (ISCO). 
\end{abstract}

\section{Introduction}

Inspiraling neutron star binaries are expected to be among the
strongest sources of gravitational radiation that could be detected by
the interferometric detectors currently under construction (GEO600,
LIGO, TAMA and Virgo).  These binary systems are therefore subject to
numerous theoretical studies.  Among them are 
(i) Post-Newtonian (PN) analytical treatments (e.g. \cite{BlancI98},
\cite{BuonaD99}, \cite{Tanig99}) and (ii)
fully relativistic
hydrodynamical treatments, pioneered by the works of Oohara and
Nakamura (see e.g.  \cite{OoharN97}) and Wilson et al.
\cite{WilsoM95,WilsoMM96}. The most recent numerical calculations,
those of Baumgarte et al.  \cite{BaumgCSST97,BaumgCSST98a} and
Marronetti et al. \cite{MarroMW98}, rely on the approximations of (i)
a quasiequilibrium state and (ii) of synchronized binaries.  Whereas
the first approximation is well justified before the innermost stable
orbit, the second one does not correspond to physical situations, since
it has been shown that the gravitational-radiation driven evolution is
too rapid for the viscous forces to synchronize the spin of each
neutron star with the orbit \cite{Kocha92,BildsC92} as they do for
ordinary stellar binaries.  Rather, the viscosity is negligible and 
the fluid velocity circulation (with respect to some inertial
frame) is conserved in these systems.  Provided that the initial spins
are not in the millisecond regime, this means that close binary
configurations are better approximated by zero vorticity (i.e. {\em
irrotational}) states than by synchronized states. 

Dynamical calculations by Wilson et al.
\cite{WilsoM95,WilsoMM96} indicate that the neutron stars may
individually collapse into a black hole prior to merger. This
unexpected result has been called into question by a number of authors
(see Ref.~\cite{MatheMW98} for a summary of all the criticisms and
their answers). Recently Flanagan \cite{Flana99} has found an error in
the analytical formulation used by Wilson et al. \cite{WilsoM95,WilsoMM96}.
This error may be responsible of the observed radial instability.
As argued by Mathews et al.~\cite{MatheMW98}, one way
to settle this crucial point is to perform computations of relativistic
irrotational configurations. We have performed recently
such computations \cite{BonazGM99}. They show no compression of the stars,
although the central density decreases much less than in the 
corotating case. In the present report, we give more details about the
results presented in Ref.~\cite{BonazGM99} and extend them to the
compactification ratio $M/R=0.17$ (the results of Ref.~\cite{BonazGM99}
have been obtained for a compactification ratio $M/R=0.14$).

\section{Analytical formulation of the problem}

\subsection{Basic assumptions} \label{s:assumpt}

We have proposed a relativistic formulation for quasiequilibrium
irrotational binaries \cite{BonazGM97b} as a generalization of the
Newtonian formulation presented in Ref.~\cite{BonazGHM92}.  The method was
based on one aspect of irrotational motion, namely the {\em
counter-rotation} (as measured in the co-orbiting frame) of the fluid
with respect to the orbital motion (see also Ref.~\cite{Asada98}).
Since then, Teukolsky~\cite{Teuko98} and Shibata~\cite{Shiba98} gave two
formulations based on the definition of irrotationality, which implies
that the specific enthalpy times the fluid 4-velocity is the gradient
of some scalar field \cite{LandaL89} ({\em potential flow}).  The three
formulations are equivalent; however the one given by Teukolsky and by
Shibata greatly simplifies the problem. Consequently we used it in the present
work.  

The irrotational hypothesis amounts to say that the co-momentum density is
the gradient of a potential
\begin{equation} \label{e:irrot}
	h \, {\bf u} = \nabla \Psi \ , 
\end{equation}
where $h$ and $\bf u$ are respectively the fluid specific enthalpy and 
fluid 4-velocity. 

\begin{figure}
\centering
\epsfig{figure=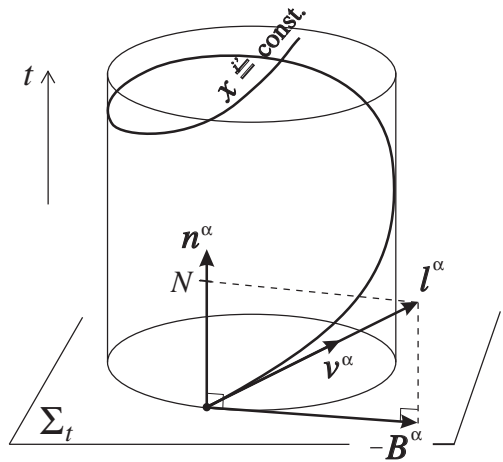,height=6cm}
\caption{Spacetime foliation $\Sigma_t$, helicoidal Killing vector
$l^\alpha$ and its trajectories $x^{i'} = {\rm const}$, which are the 
worldlines of the co-orbiting observer (4-velocity: $v^\alpha$).
Also shown are the rotating-coordinate shift vector $B^\alpha$
and the unit future-directed vector $n^\alpha$, normal to the spacelike
hypersurface $\Sigma_t$ (Figure from Ref.~\cite{BonazGM97b}).}\label{f:helic}
\end{figure}

Beside the physical assumption (\ref{e:irrot}), two simplifying approximations
are introduced:
\begin{enumerate}
\item The spacetime is supposed to have a helicoidal symmetry 
\cite{BonazGM97b}, which means
that the orbits are exactly circular and that the gravitational radiation
content of spacetime is neglected. 
\item The spatial 3-metric is assumed to be conformally flat (Wilson-Mathews
approximation \cite{WilsoM89,WilsoM95}), so that the full spacetime 
metric reads
\begin{equation} \label{e:met}
ds^2 = -(N^2 - B_i B^i) dt^2 - 2 B_i dt\, dx^i 
        + A^2 f_{ij} dx^i\, dx^j \ ,
\end{equation}
where $f_{ij}$ is the flat space metric.
\end{enumerate}

The Killing vector corresponding to hypothesis (i) is
denoted by $\bf l$ (cf. Fig.~\ref{f:helic}).

Approximation (i) is physically well motivated, at least up to the
innermost stable orbit. Regarding the second approximation, it is to be 
noticed that (i) the 1-Post Newtonian (PN) approximation
to Einstein equations fits this form, (ii) it is exact for arbitrary
relativistic spherical configurations and (iii) it is very accurate for
axisymmetric rotating neutron stars \cite{CookST96}.  An interesting discussion
about some justifications of the Wilson-Mathews approximation may be
found in \cite{MatheMW98}. A stronger justification may be obtained by
considering the 2.5-PN metric obtained by Blanchet et al. 
\cite{BlancFP98} for point mass binaries. Using Eq.~(7.6) of 
Ref.~\cite{BlancFP98}, the deviation from a conformally flat 3-metric
(which occurs at the 2-PN order) can be computed at the location of one
point mass (i.e. where it is expected to be maximal),
the 3-metric $h_{ij}$ being written as
\begin{equation} \label{e:met-PN}
   h_{ij} = A^2 \, f_{ij} + h_{ij}^{\rm 2PN} + h_{ij}^{\rm 2.5PN} \ .
\end{equation}
The result is shown in Table~\ref{t:PN} for two stars of $1.4 M_\odot$ each.
It appears that at a separation as close as $30{\ \rm km}$, where the two
stars certainly almost touch each other, the relative deviation from a 
conformally flat 3-metric is below $2\%$. 

\begin{table}[t]
\begin{center}
\begin{tabular}{*{6}{c}}
\hline
\\[0.5ex]
$d$ & $v/c$ & $\omega_{\rm 2PN}$ & Conformal fact. 
	 & $\left| h_{ij}^{\rm 2PN} \right| /A^2$ 
         & $\left| h_{ij}^{\rm 2.5PN} \right| /A^2$ \\
$\rm [km]$ &  & [rad/s] & $A^2$ & & \\[0.5ex]
\hline
\\[0.5ex]
100 & 0.10 & 579 & 1.04 & $1.8\times 10^{-3}$ & $6.3\times 10^{-4}$
\\[0.5ex]
50  & 0.13  & 1572 & 1.09 & $6.8\times 10^{-3}$ & $3.3\times 10^{-3}$
\\[0.5ex]
40  & 0.14  & 2166 & 1.11 & $1.0\times 10^{-2}$ & $5.6\times 10^{-3}$
\\[0.5ex]
30 & 0.16 & 3387 & 1.15 & $1.8\times 10^{-2}$  & $1.1\times 10^{-2}$
\\[0.5ex]
\hline
\end{tabular}
\vspace{3mm}
\caption{Deviation from a conformally flat 3-metric at 2-PN at 2.5-PN order
for point mass binaries of $M=1.4\, M_\odot$ each. $d$ is the coordinate
separation between the two stars, $v$ is the coordinate orbital velocity
of one star, $\omega_{\rm 2PN}$ is the orbital angular velocity, the other
notations are defined by Eq.~(\ref{e:met-PN}). The metric at the
2.5PN level is taken from Blanchet et al.~\cite{BlancFP98} and is computed
at the location of the point masses.}
\label{t:PN}
\end{center}
\end{table}

\subsection{Equations to be solved} 

We refer to Ref.~\cite{BonazGM99} for the presentation of the partial
differential equations (PDEs) which result from the assumptions presented above.
In the present report, let us simply mention a point which seems to have
been missed by various authors: the existence of the first integral of motion
\begin{equation} \label{e:int_prem}
   h \, {\bf l} \cdot {\bf u} = {\rm const.} 
\end{equation}
does not result solely from the existence of the helicoidal 
Killing vector $\bf l$. Indeed, Eq.~(\ref{e:int_prem})
is not merely the relativistic generalization of the Bernoulli
theorem which states that $h \, {\bf l} \cdot {\bf u}$ is constant along
each single streamline and which results directly from the existence of
a Killing vector without any hypothesis on the flow. In order for the
constant to be uniform over the streamlines (i.e. to be a constant over
spacetime), as in Eq.~(\ref{e:int_prem}), some additional property of the
flow must be required. One well known possibility is {\em rigidity} (i.e.
$\bf u$ colinear to $\bf l$) \cite{Boyer65}. The alternative property
with which we are concerned here is {\em irrotationality}
[Eq.~(\ref{e:irrot})]. This was first pointed out by Carter
\cite{Carte79}.

\begin{figure}
\centering
\epsfig{figure=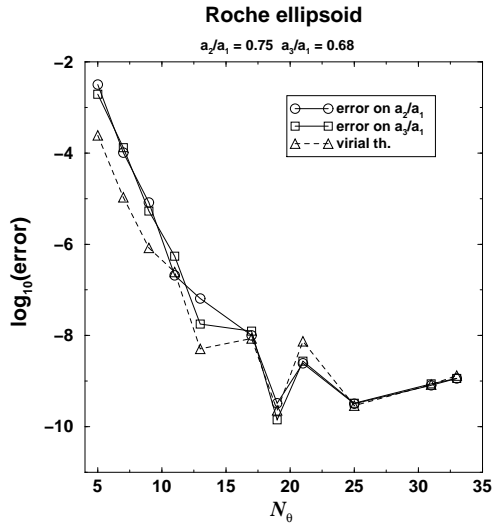,height=7cm}
\caption{Logarithm of the relative global error of the numerical solution
with respect to the number of degrees of freedom in
$\theta$ for a Roche ellipsoid for an equal mass binary system and
$\Omega^2/(\pi G\rho) = 0.1147$ (the numbers of degrees of freedom in the
other directions are $N_r = 2N_\theta-1$ and $N_\varphi = N_\theta -1$). 
Also shown is the error in the
verification of the virial theorem (Figure from Ref.~\cite{BonazGM98a}).}
\label{f:test:roche}
\end{figure}

\section{Numerical procedure}

\subsection{Description}

The numerical procedure to solve the PDE system
is based on the multi-domain spectral method presented in Ref.~\cite{BonazGM98a}.
We simply recall here some basic features of the method:
\begin{itemize}
\item Spherical-type coordinates $(\xi,\theta',\varphi')$ centered on each 
star are used: this ensures
a much better description of the stars  than with Cartesian coordinates. 
\item These spherical-type coordinates are surface-fitted coordinates: i.e.
the surface of each star lies at a constant value of the coordinate
$\xi$ thanks to a mapping $(\xi,\theta',\varphi')\mapsto (r,\theta,\varphi)$
(see  \cite{BonazGM98a} for details about this mapping). This 
ensures that the spectral method applied in each domain is free from any
Gibbs phenomenon. 
\item The outermost domain extends up to spatial infinity, thanks to the
mapping $1/r = (1-\xi)/(2R_0)$. This enables to put exact boundary conditions
on the elliptic equations for the metric coefficients: spatial infinity
is the only location where the metric is known in advance (Minkowski metric). 
\item Thanks to the use of a spectral method \cite{BonazGM99b}
in each domain, the numerical
error is {\em evanescent}, i.e. it decreases exponentially with the number
of coefficients (or equivalently grid points) used in the spectral expansions,
as shown in Fig.~\ref{f:test:roche}.
\end{itemize}

\begin{figure}
\centering
\epsfig{figure=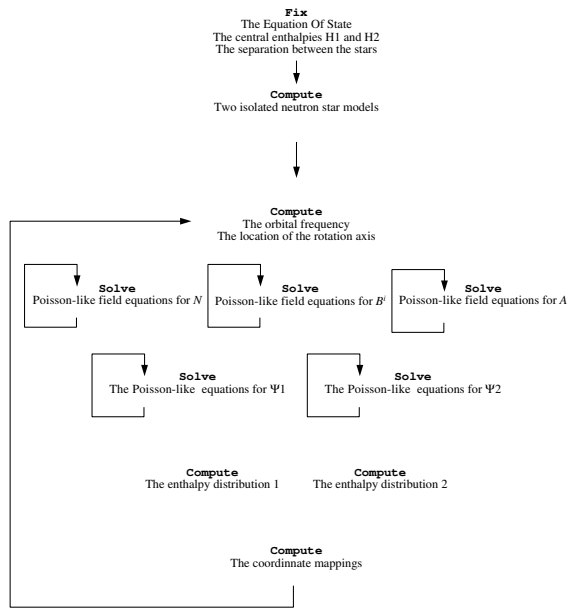,height=8cm}
\caption{Schematic representation of the iterative procedure used in the
numerical code.}
\label{f:iter}
\end{figure}

\begin{figure}
\centering
\epsfig{figure=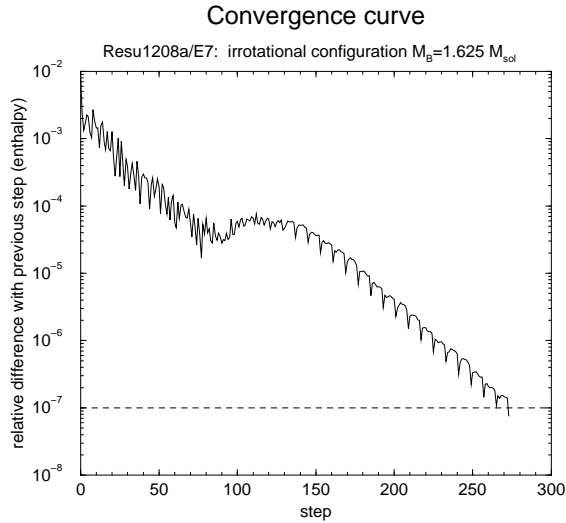,height=7cm}
\caption{Convergence (measured by the relative difference in the enthalpy field
between two successive steps) of the iterative procedure for one of the
irrotational models considered in Ref.~\cite{BonazGM99}. The bump around
the 90th step corresponds to the switch on of the procedure of convergence 
toward a given baryon mass.}
\label{f:converge}
\end{figure}

The PDE system to be solved being non-linear, we use an iterative 
procedure. This procedure is sketched in Fig.~\ref{f:iter}.
The iteration is stopped when the relative difference in the enthalpy field
between two successive steps goes below a certain threshold, typically
$10^{-7}$ (cf. Fig.~\ref{f:converge}). 

The numerical code is written in the {\sc LORENE} language 
\cite{MarckG99}, which is a C++ based language for numerical relativity. 
A typical run make use of $N_r = 33$, $N_\theta = 21$, and
$N_\varphi=20$ coefficients (= number of
collocation points, which may be seen as number of grid points) in
each of the domains on the multi-domain spectral method. 8 domains are
used : 3 for each star and 2 domains centered on the intersection between
the rotation axis and the orbital plane. The corresponding memory requirement
is 155 MB. A computation involves $\sim 250$ steps (cf. Fig.~\ref{f:converge}),
which takes 9~h~30~min on one CPU of a SGI Origin200 computer 
(MIPS~R10000 processor at 180 MHz). 
Note that due to the rather small memory requirement, 
runs can be performed in parallel on a multi-processor platform. 
This especially useful to compute sequences of configurations.  

\subsection{Tests passed by the code}

In the Newtonian and incompressible
limit, the analytical solution constituted by a Roche ellipsoid is
recovered with a relative accuracy of $\sim 10^{-9}$, as shown in  
Fig.~\ref{f:test:roche}. For compressible and irrotational Newtonian
binaries, no analytical solution is available, but the virial theorem
can be used to get an estimation of the numerical error: we found that the
virial theorem is satisfied with a relative accuracy of $10^{-7}$. A
detailed comparison with the irrotational Newtonian configurations
recently computed by Uryu \& Eriguchi \cite{UryuE98a,UryuE98b} will be
presented elsewhere.  Regarding the relativistic case, we
have checked our procedure of resolution of the gravitational field
equations by comparison with the results of Baumgarte et
al.~\cite{BaumgCSST98a} which deal with corotating binaries [our code
can compute corotating configurations by setting to zero the
velocity field of the fluid with respect to the co-orbiting observer]. 
We have performed the comparison with
the configuration $z_A=0.20$ in Table~V of Ref.~\cite{BaumgCSST98a}. We
used the same equation of state (EOS) (polytrope with $\gamma=2$), same
value of the separation $r_C$ and same value of the maximum density parameter
$q^{\rm max}$. We found a relative discrepancy of $1.1\%$ on $\Omega$,
$1.4\%$ on $M_0$, $1.1\%$ on $M$, $2.3\%$ on $J$, $0.8\%$ on $z_A$,
$0.4\%$ on $r_A$ and $0.07\%$ on $r_B$ (using the notations of
Ref.~\cite{BaumgCSST98a}).

\section{Numerical results} \label{s:num_res}

\subsection{Equation of state and compactification ratio}

As a model for nuclear matter, 
we consider a polytropic equation of state (EOS) with an adiabatic index
$\gamma=2$: 
\begin{equation} \label{e:eos}
p=\kappa (m_{\rm B} n)^\gamma \ , \qquad 
e=m_{\rm B} n + p/(\gamma-1) \ ,
\end{equation}
where $p$, $n$, $e$ are respectively the fluid pressure, baryon density and
proper energy density, and $m_{\rm B} = 1.66\times 10^{-27} {\rm\ kg}$, 
$\kappa = 1.8 \times 10^{-2} {\ \rm J\, m}^3{\rm kg}^{-2}$. This EOS is the
same as that used by Mathews, Marronetti and Wilson
(Sect.~IV~A of Ref~\cite{MatheMW98}).

\begin{figure}
\centering
\epsfig{figure=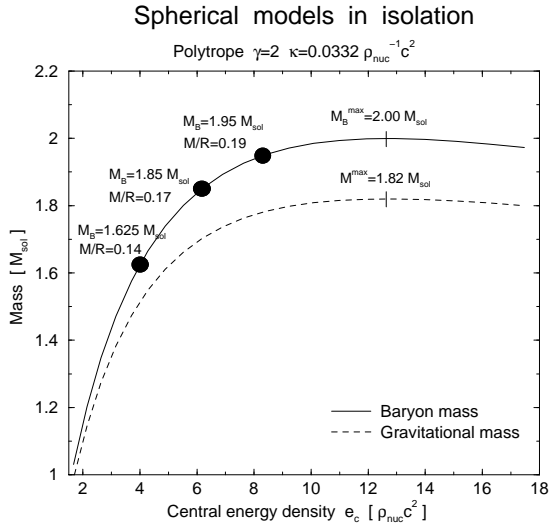,height=7cm}
\caption{Mass as a function of the central energy density for
static isolated neutron stars constructed with the EOS (\ref{e:eos}).
The heavy dots are configurations considered by our group 
\cite{BonazGM99} and
Marronetti, Mathews \& Wilson group \cite{MatheMW98}, \cite{MarroMW99}
(see text)
($\rho_{\rm nuc} := 1.66\times 10^{17} {\rm \ kg\, m}^{-3}$).}
\label{f:m_sher}
\end{figure}

The mass -- central density curve of static
configurations in isolation constructed upon this EOS is represented
in Fig.~\ref{f:m_sher}. The three points on this curve corresponds to 
three configurations studied by our group and that of Marronetti, Mathews
and Wilson:
\begin{itemize}
\item The configuration of baryon mass $M_{\rm B} = 1.625 \, M_\odot$ 
and compactification ratio $M/R = 0.14$ is
that considered in the dynamical study of Mathews, Marronetti and Wilson
\cite{MatheMW98} and in the quasiequilibrium study of our group 
(Ref.~\cite{BonazGM99} and this paper). 
\item The configuration of baryon mass $M_{\rm B} = 1.85 \, M_\odot$ 
and compactification ratio $M/R = 0.17$ is studied in the present paper.
\item The configuration of baryon mass $M_{\rm B} = 1.95 \, M_\odot$ 
and compactification ratio $M/R = 0.19$ has been studied recently by
Marronetti, Mathews and Wilson \cite{MarroMW99}\footnote{Marronetti et al.
\cite{MarroMW99} use a different value for the EOS constant $\kappa$:
their baryon mass $M_{\rm B} = 1.55 \, M_\odot$ must be rescaled to our
value of $\kappa$ in order to get $M_{\rm B} = 1.95 \, M_\odot$. Apart from
this scaling, this is the same configuration, i.e. it has the same 
compactification ratio $M/R = 0.19$ and its relative distance with respect to 
the maximum mass configuration, as shown in Fig.~\ref{f:m_sher}, is the same.}
by means of a new code for quasiequilibrium irrotational configurations.
\end{itemize}

\subsection{Irrotational sequence with $M/R=0.14$}

\begin{figure}
\centering
\epsfig{figure=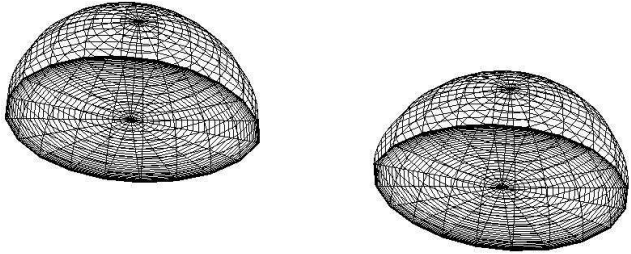,height=7cm}
\caption{Shape of irrotational binary neutron stars of baryon mass 
$M_{\rm B} = 1.625 \, M_\odot$, when the coordinate separation between
their centers (density maxima) is $41{\rm\ km}$. Only one half of each star
is represented (the part which is above the orbital plane).
The drawing is that of the numerical grid, which coincides with the surface
of the star, thanks to the use of surface-fitted spherical coordinates.}
\label{f:surf-3D}
\end{figure}

In this section, we give some details about the irrotational sequence
$M_{\rm B} = 1.625 \, M_\odot$ presented in Ref.~\cite{BonazGM99}.
This sequence starts at the coordinate separation $d=110 {\rm\ km}$ (orbital
frequency $f=82{\rm\ Hz}$), where the two stars are almost spherical, and
ends at $d=41 {\rm\ km}$ ($f=332{\rm\ Hz}$), where a cusp appears on the
surface of the stars, which means that the stars start to break. 
The shape of the surface at this last point is shown in Fig.~\ref{f:surf-3D}.

\begin{figure}
\centering
\epsfig{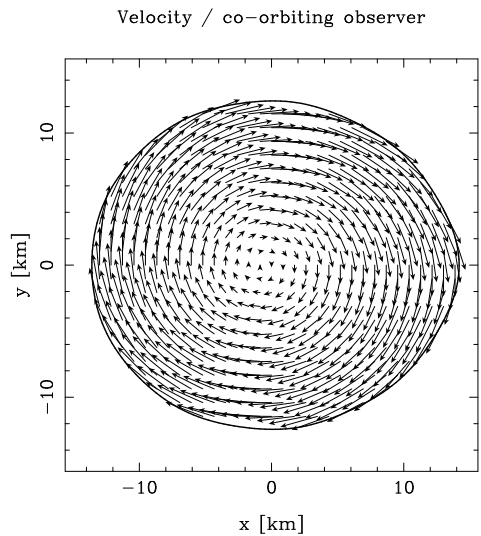}
\caption{Velocity field with respect to the co-orbiting observer, for the
configuration shown in Fig.~\ref{f:surf-3D}. The plane of the figure is the
orbital plane. The heavy line denotes the surface of the star.
The companion is located at $x=+41{\rm\ km}$.}
\label{f:vit}
\end{figure}

The velocity field with respect to the co-orbiting observer, as defined 
by Eq.~(52) of Ref.~\cite{BonazGM97b}, is shown in Fig.~\ref{f:vit}. 
Note that this field is tangent to the surface of the star, as it must be.

\begin{figure}
\centering
\epsfig{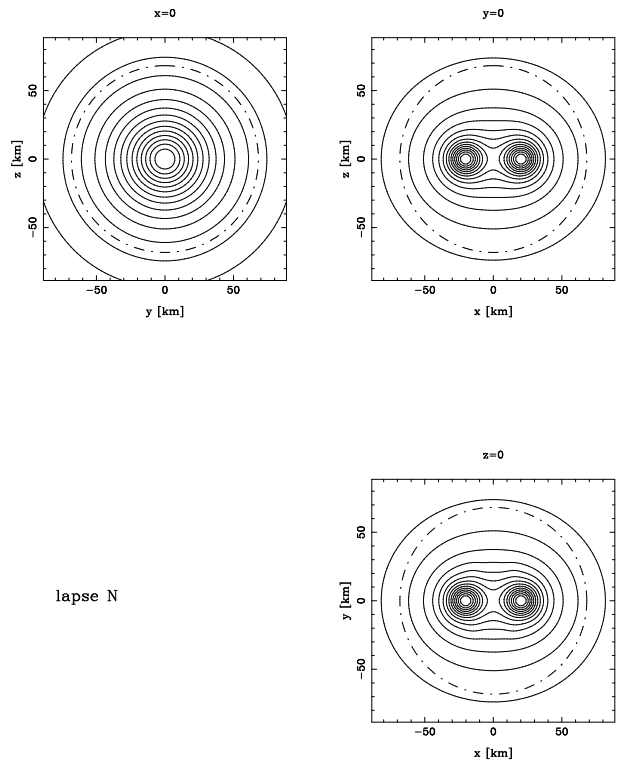}
\caption{Isocontour of the lapse function $N$ for the
configuration shown in Fig.~\ref{f:surf-3D}. The plots are cross section
is the $x=0$, $y=0$ and $z=0$ planes (note that the $x$ coordinate is
shifted by $20.5{\rm\ km}$ with respect to that of Fig.~\ref{f:vit}).
The dot-dashed line denotes the
boundary between the inner numerical grid and the outer compactified one 
(which extends to spatial infinity), for the grid system centered on
the intersection between the rotation axis and the orbital plane.}
\label{f:lapse}
\end{figure}

The lapse function $N$ (cf. Eq.~\ref{e:met}) is represented in 
Fig.~\ref{f:lapse}. The coordinate system $(x,y,z)$ is centered on
the intersection between the rotation axis and the orbital plane.
The $x$ axis joins the two stellar centers, and the orbital is the $z=0$
plane. The value of $N$ at the center of each star is
$N_{\rm c}=0.64$. 

\begin{figure}
\centering
\epsfig{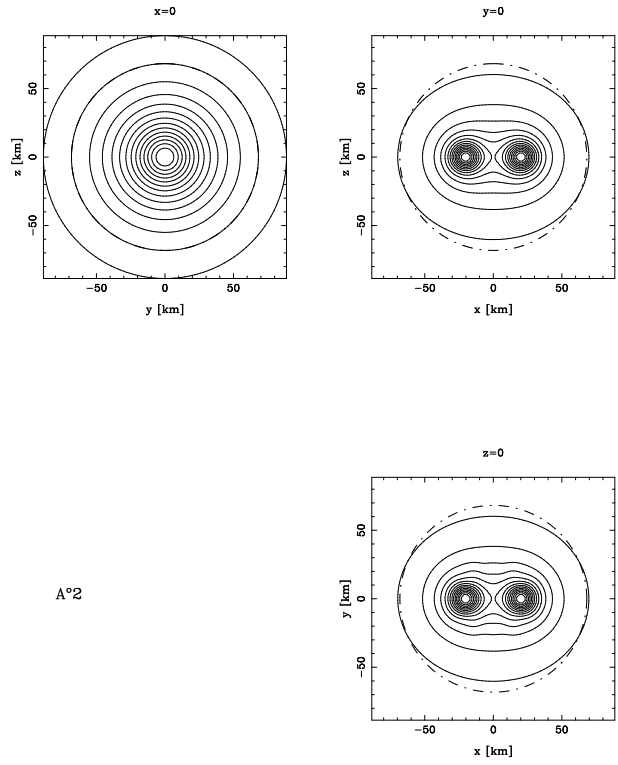}
\caption{Same as Fig.~\ref{f:lapse} but for the conformal factor $A^2$
of the spatial metric.}
\label{f:acar}
\end{figure}

The conformal factor $A^2$ of the 3-metric [cf. Eq.~(\ref{e:met})] is
represented in Fig.~\ref{f:acar}. Its value at the center of each star is
$A^2_{\rm c}=2.20$. 

\begin{figure}
\centering
\epsfig{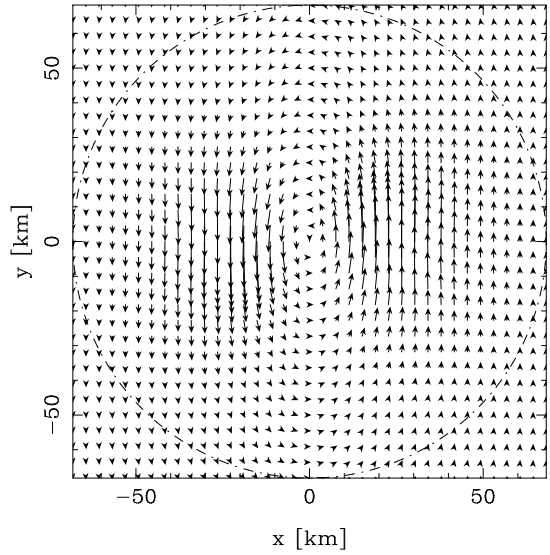}
\caption{Shift vector of nonrotating coordinates in the orbital plane, 
for the configuration shown in Figs.~\ref{f:surf-3D}-\ref{f:acar}.}
\label{f:shift}
\end{figure}

The shift vector of nonrotating coordinates, $\bf N$, (defined by Eq.~(9) of 
Ref.~\cite{BonazGM99}) is shown in Fig.~\ref{f:shift}.
Its maximum value is $0.10\, c$. 

\begin{figure}
\centering
\epsfig{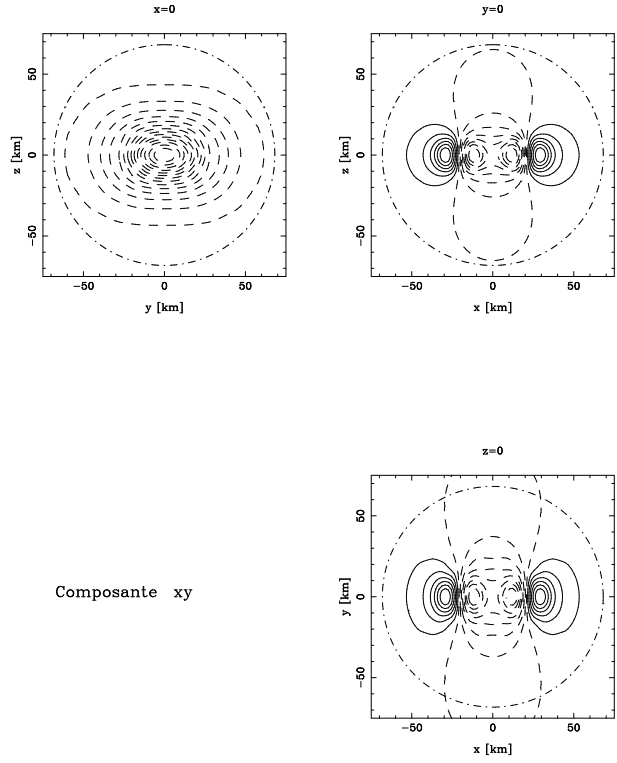}
\caption{Same as Fig.~\ref{f:lapse} but for the component $K_{xy}$ of
the extrinsic curvature tensor. The solid (resp. dashed) lines corresponds
to positive (resp. negative) values of $K_{xy}$.}
\label{f:kxy}
\end{figure}

The $K_{xy}$ component of the extrinsic curvature tensor of the hypersurfaces
$t={\rm const}$ is shown in Fig.~\ref{f:kxy}. We chose to represent the
$K_{xy}$ component because it is the only component of $K_{ij}$ for which
none of the sections in the three planes $x=0$, $y=0$ and $z=0$
vanishes.

\begin{figure}
\centering
\epsfig{figure=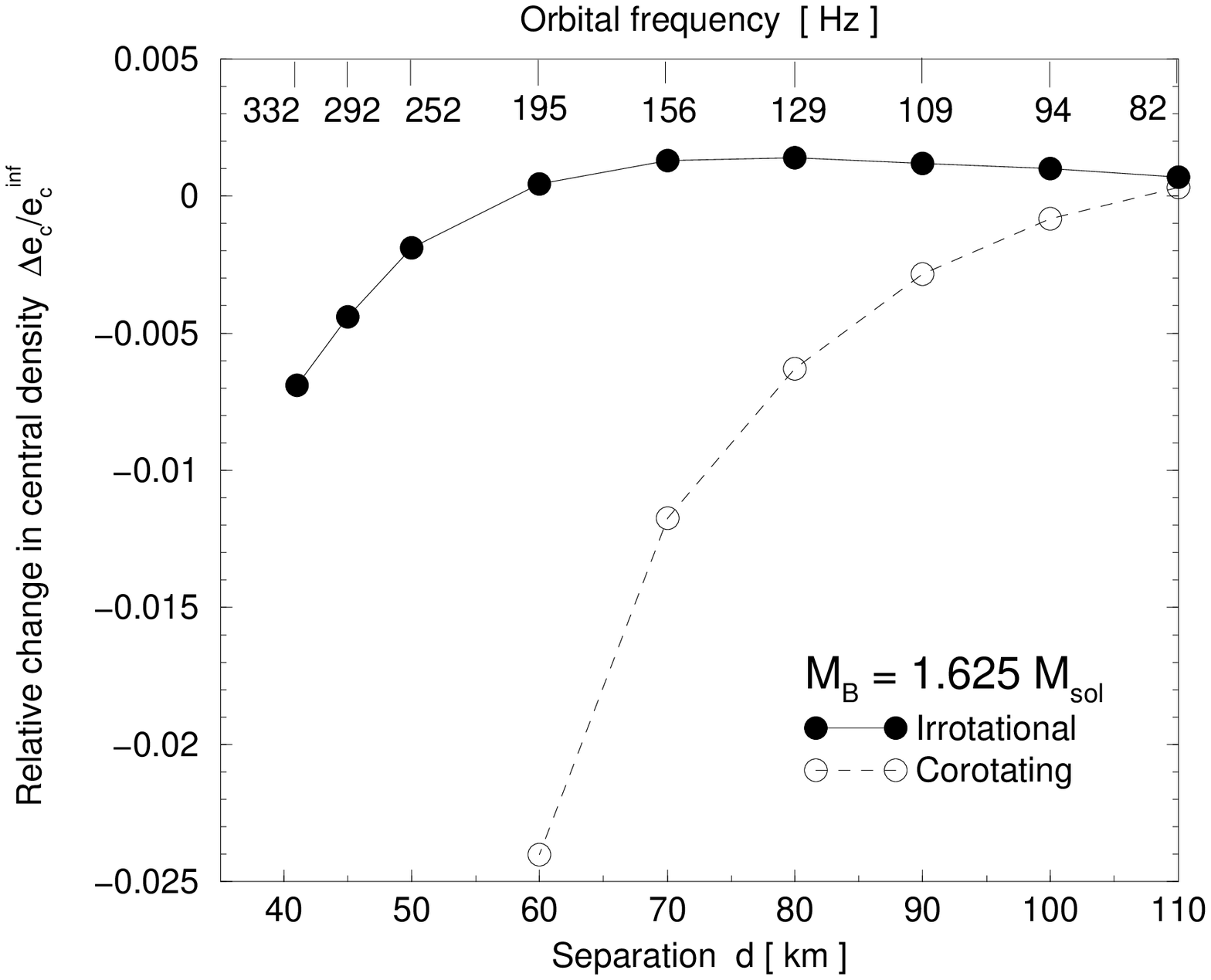,height=7cm}
\caption{Relative variation of the central energy density $e_{\rm c}$ with
respect to its value at infinite separation $e_{\rm c}^{\rm inf}$ as a
function of the coordinate separation $d$ (or of the orbital frequency
$\Omega/(2\pi)$) for constant baryon mass $M_{\rm B} = 1.625 \,
M_\odot$ sequences.  The solid (resp. dashed) line corresponds to a
irrotational (resp. corotating) sequence of coalescing neutron star
binaries (Figure from Ref.~\cite{BonazGM99}).}
\label{f:cent_dens_1.625}
\end{figure}

The variation of the central density along the $M_{\rm B}=1.625\, M_\odot$
sequence is shown in Fig.~\ref{f:cent_dens_1.625}. We have also
computed a corotating sequence for comparison (dashed line in 
Fig.~\ref{f:cent_dens_1.625}). In the corotating case, the central density
decreases quite substantially as the two stars approach each other. 
This is in agreement with the results of Baumgarte et 
al.~\cite{BaumgCSST97,BaumgCSST98a}. In the irrotational case (solid
line in Fig.~\ref{f:cent_dens_1.625}), the central density remains rather
constant (with a slight increase, below $0.1\%$)
before decreasing. We can thus conclude that no tendency to individual
gravitational collapse is found in this case. 
This contrasts with results of dynamical calculations by 
Mathews et al.~\cite{MatheMW98} which show a central density increase
of $14\%$ for the same compactification ratio $M/R=0.14$.

\subsection{Irrotational sequence with $M/R=0.17$}

\begin{figure}
\centering
\epsfig{figure=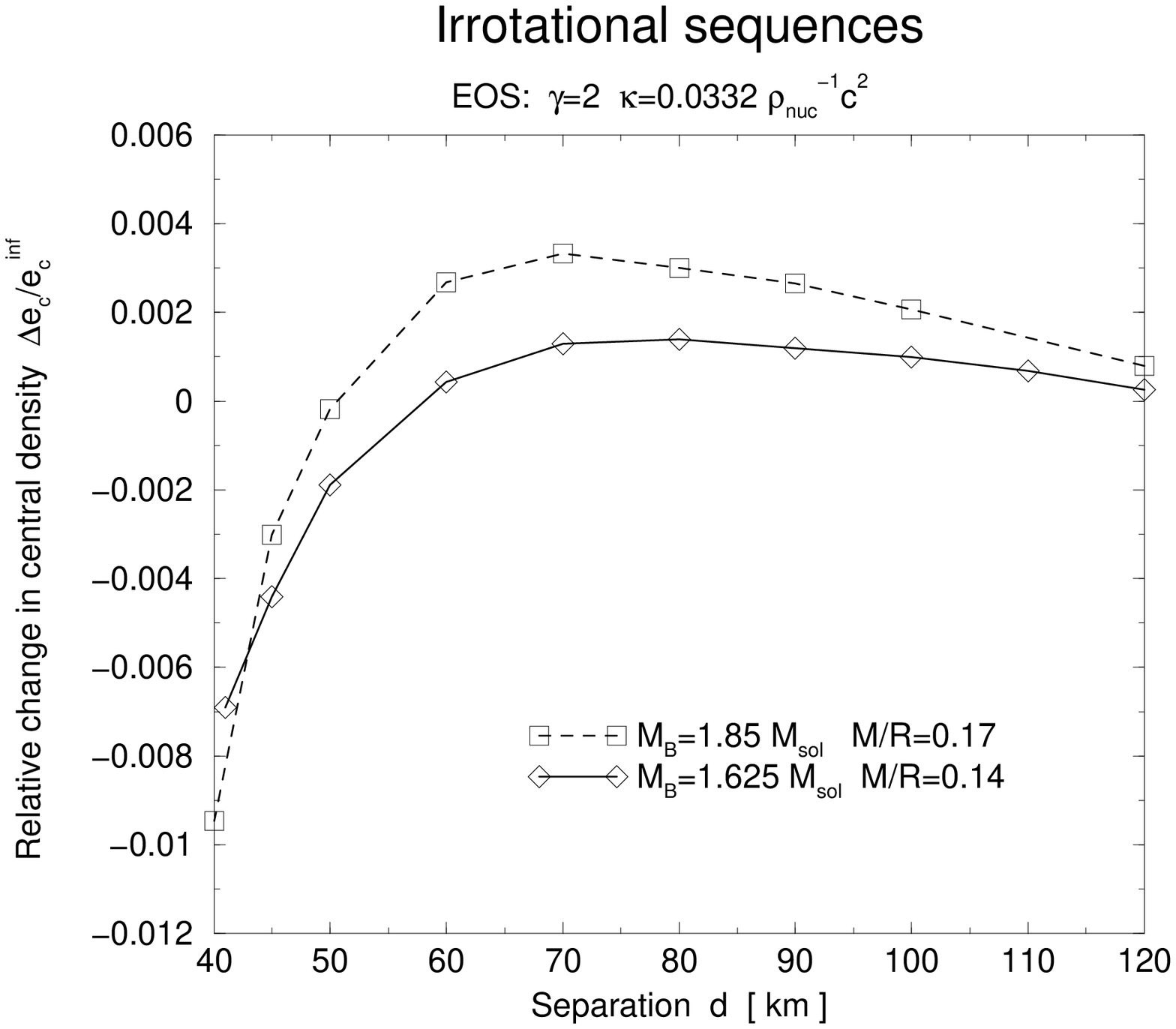,height=7cm}
\caption{Relative variation of the central energy density $e_{\rm c}$ with
respect to its value at infinite separation $e_{\rm c}^{\rm inf}$ as a
function of the coordinate separation $d$ for constant baryon mass 
sequences with 
$M_{\rm B} = 1.625 \, M_\odot$ (solid line, same as in 
Fig.~\ref{f:cent_dens_1.625}) and
$M_{\rm B} = 1.85\, M_\odot$ (dashed line).}
\label{f:cent_dens_1.625_1.85}
\end{figure}

In order to investigate how the above result depends on the compactness
of the stars, we have computed an irrotational sequence with a baryon
mass $M_{\rm B} = 1.85\, M_\odot$, which corresponds to a compactification
ratio $M/R=0.17$ for stars at infinite separation
(second heavy dot in Fig.~\ref{f:m_sher}). 
The result is compared
with that of $M/R=0.14$ in Fig.~\ref{f:cent_dens_1.625_1.85}. 
A very small density increase (at most $0.3\%$) 
is observed before the decrease. Note that this density increase remains
within the expected error ($\sim 2\%$, cf. Sect.~\ref{s:assumpt}) induced
by the conformally flat approximation for the 3-metric, so that it cannot
be asserted that this effect would remain in a complete calculation.

Marronetti, Mathews and Wilson~\cite{MarroMW99} have recently computed
quasiequilibrium irrotational configurations by means of a new code.
They use a higher 
compactification ratio, $M/R=0.19$ (third heavy dot in Fig.~\ref{f:m_sher}).
They found a central density increase as the orbit shrinks much
pronounced than that we found for the compactification ratio $M/R=0.17$:
$3.5\%$ against $0.3\%$. We will present irrotational sequences with
the compactification ratio $M/R=0.19$ and compare with the results by
Marronetti et al.~\cite{MarroMW99} in a future article. 
 
\section{Innermost stable circular orbit}

An important parameter for the detection of a gravitational wave signal from
coalescing binaries is the location of the innermost stable circular 
orbit (ISCO), if any. In Table~\ref{t:isco}, we recall what is known about
the existence of an ISCO for extended fluid bodies. The case of two point
masses is discussed in details in Ref.~\cite{BuonaD99}.

\begin{table}[t]
\begin{center}
\begin{tabular}{*{3}{c}}
\hline
\\[0.5ex]
Model & Existence of an ISCO & References \\[0.5ex]
\hline
\\[0.5ex]
Newtonian corotating & ISCO $\Leftrightarrow \, \gamma > 2$ & \cite{ShibaTN97}
\\[0.5ex]
Newtonian irrotational & ISCO $\Leftrightarrow \, \gamma > 2.4$ &
\cite{UryuE98a} 
\\[0.5ex]
GR corotating & ISCO $\Leftrightarrow \, \gamma > 5/3$ &
\cite{BaumgCSST97}, \cite{ShibaTN97} 
\\[0.5ex]
GR irrotational & ISCO exists for $\gamma = \infty$ &
\cite{Tanig99} 
\\[0.5ex]
\hline
\end{tabular}
\vspace{3mm}
\caption{Known results about the existence of an ISCO for extended fluid
bodies, in terms of the adiabatic index $\gamma$.}
\label{t:isco}
\end{center}
\end{table}

\begin{figure}
\centering
\epsfig{figure=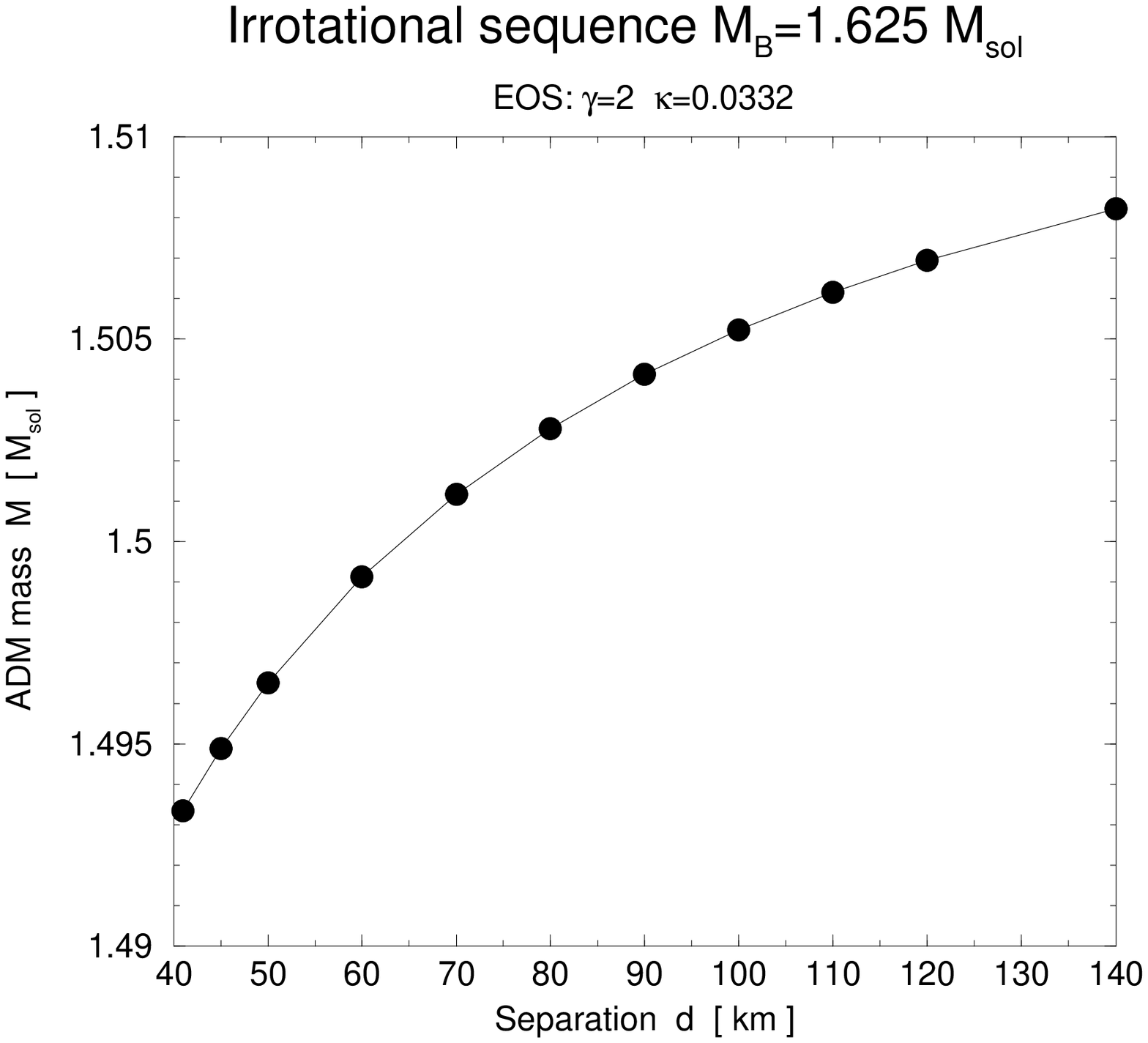,height=6.5cm}
\caption{Half of the ADM mass of the binary system as a function of the
coordinate distance $d$, along the evolutionnary sequence 
$M_{\rm B} = 1.625 \, M_\odot$.}
\label{f:ADM_1.625}
\end{figure}

\begin{figure}
\centering
\epsfig{figure=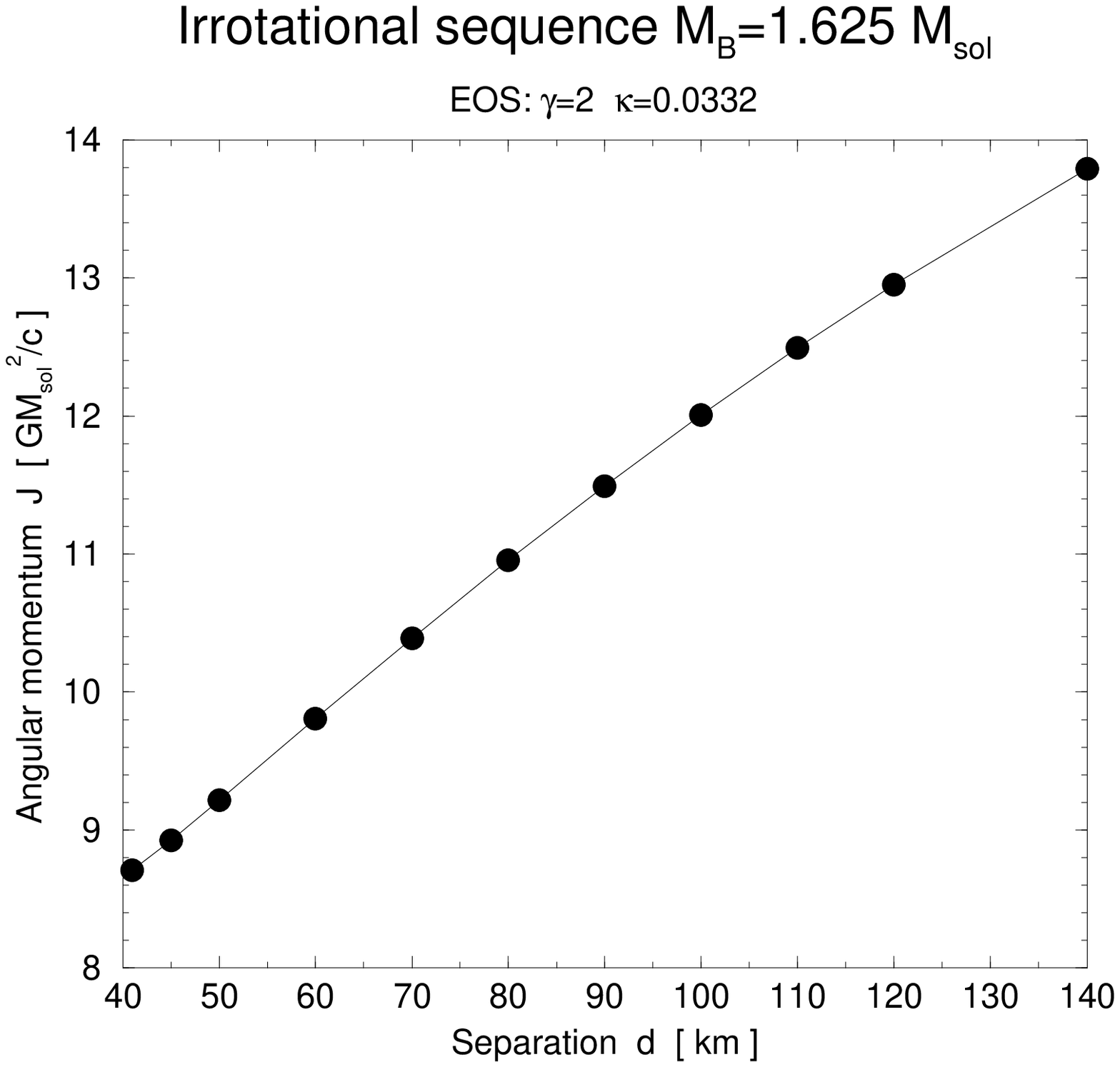,height=6.5cm}
\caption{Total angular momentum of the binary system as a function of the
coordinate distance $d$, along the evolutionary sequence 
$M_{\rm B} = 1.625 \, M_\odot$.}
\label{f:J_1.625}
\end{figure}

\begin{figure}
\centering
\epsfig{figure=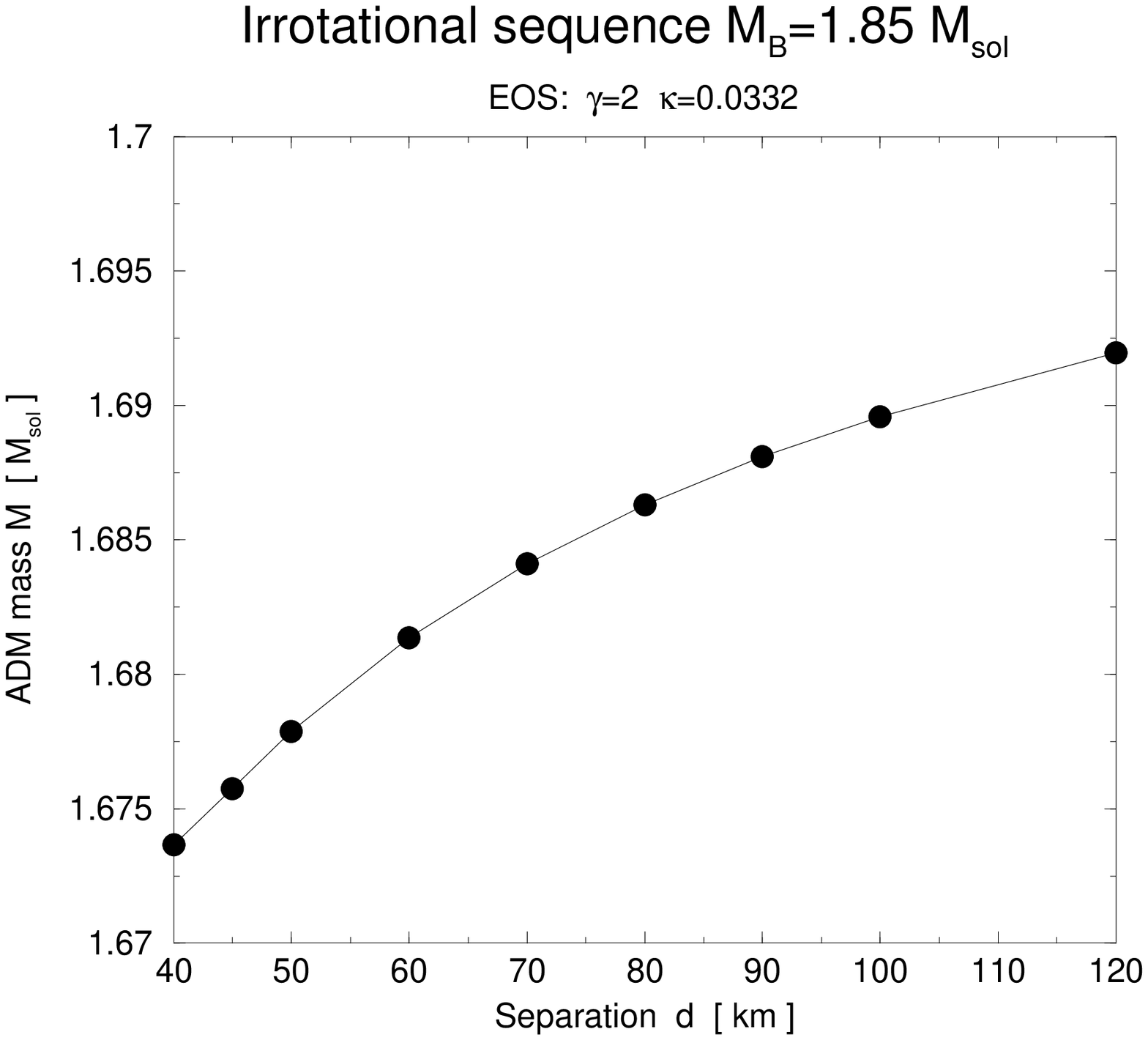,height=6.5cm}
\caption{Half of the ADM mass of the binary system as a function of the
coordinate distance $d$, along the evolutionary sequence 
$M_{\rm B} = 1.85 \, M_\odot$.}
\label{f:ADM_1.85}
\end{figure}

\begin{figure}
\centering
\epsfig{figure=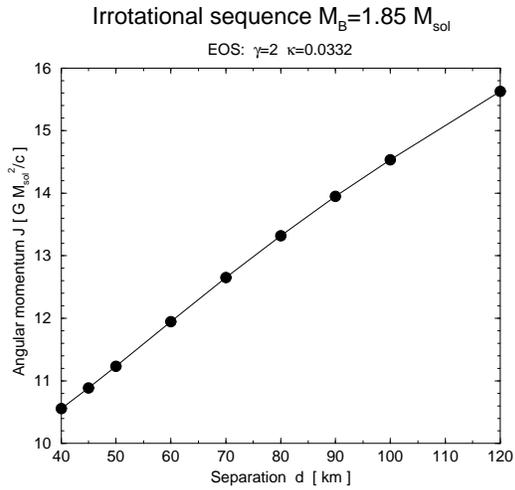,height=6.5cm}
\caption{Total angular momentum of the binary system as a function of the
coordinate distance $d$, along the evolutionary sequence 
$M_{\rm B} = 1.85 \, M_\odot$.}
\label{f:J_1.85}
\end{figure}

For Newtonian binaries, it has been shown \cite{LaiRS93} that the ISCO
is located at a minimum of the total energy, as well as total angular momentum,
along a sequence at constant baryon number and constant circulation 
(irrotational sequences are such sequences). The instability found in this
way is dynamical. For corotating sequences, it is secular instead 
\cite{LaiRS93,LaiRS94}.
This turning point method also holds for locating ISCO in 
relativistic corotating binaries \cite{BaumgCSST98b}. 
For relativistic irrotational configurations, no rigorous theorem has been
proven yet about the localization of the ISCO by a turning point method. 
All what can be said is that no turning point is present in the 
irrotational sequences considered in Sect.~\ref{s:num_res}:
Fig.~\ref{f:ADM_1.625} shows the variation as the orbit shrinks 
of the ADM mass of the spatial hypersurface $t={\rm const}$ (which is a 
measure of the total energy, or equivalently of the
binding energy, of the system) for the $M_{\rm B}=1.625\, M_\odot$
sequence. Clearly, the ADM mass decreases
monoticaly, without showing any turning point. Figure~\ref{f:J_1.625}
shows the evolution of the total angular momentum along the same sequence.
Again there is no turning point. 
The same behaviour holds for the $M_{\rm B}=1.85\, M_\odot$ sequence, as
shown in Figs.~\ref{f:ADM_1.85} and \ref{f:J_1.85}.

\section{Conclusion and perspectives}

We have computed evolutionary sequences of quasiequilibrium irrotational
configurations of binary stars in general relativity. 
The evolution of the central density of each star have been monitored 
as the orbit shrinks.
For a compactification ratio $M/R=0.14$, the central density 
remains rather constant (with a slight increase, below $0.1\%$)
before decreasing. For a higher compactification ratio $M/R=0.17$ (i.e. stars
closer to the maximum mass configuration), a very small density increase
(at most $0.3\%$) is observed before the density 
decrease.
It can be thus concluded
that no substantial compression of the stars is found, which means
that no tendency to individually collapse to black hole prior to merger
is observed. 
Moreover, the observed density increase remains
within the expected error ($\sim 2\%$, cf. Sect.~\ref{s:assumpt}) induced
by the conformally flat approximation for the 3-metric, so that it cannot
be asserted that this effect would remain in a complete calculation.

No turning point has been found in the binding energy or angular
momentum along evolutionary sequences, which may indicate that these
systems do not have any innermost stable circular orbit (ISCO). 

All these results have been obtained for a polytropic EOS with the
adiabatic index $\gamma=2$. We plan to extend them to other values of
$\gamma$ in the near future. We also plan to abandon the conformally
flat approximation for the 3-metric and use the full Einstein equations,
keeping the helicoidal symmetry in a first stage.

\section*{Acknowledgments}

We would like to thank Jean-Pierre Lasota for his constant support and 
Brandon Carter for illuminating discussions. 
The numerical calculations have been performed on computers 
purchased thanks to a special grant from the SPM and SDU departments of CNRS.

\end{document}